\documentclass[aps,prl,reprint]{revtex4-2}

\usepackage[utf8]{inputenc}
\usepackage{amsthm}
\usepackage{amsmath}
\usepackage{color}
\usepackage{graphicx,xcolor}
\usepackage{amssymb}
\usepackage{bbold}
\usepackage{gensymb}
\usepackage{tikz}

\usepackage{amsmath,amssymb,bm}
\usepackage{xcolor}
\usepackage[colorlinks=true,linkcolor=blue!50!black,citecolor=blue!50!black,urlcolor=blue!50!black]{hyperref}

\newcommand{\dt}{\delta t}

\newcommand{\ii}{\mathrm{i}} 
\newcommand{\eul}{\mathrm{e}} 
\newcommand{\diff}{\mathrm{d}} 
\newcommand{\id}{\mathbb{1}} 
\newcommand{\braket}[1]{\langle#1\rangle} 

\begin{document}

\title{The TEMPO network can be made independent of the system size}

\author{Valentin Link}
\affiliation{Institut für Physik und Astronomie, Technische Universität Berlin, D-10623, Berlin, Germany}

\begin{abstract}
The time-evolving matrix product operator (TEMPO) method efficiently simulates non-Markovian open quantum system dynamics by contracting a two-dimensional tensor network representing the time-discretized Feynman–Vernon influence functional. The standard form of this network has an unfavorable scaling with the size of the system. We show that, within the usual second-order time-step error, the dimensions of all internal network bonds can be reduced to $(2m+1)(m+1)$, where $m$ is the number of independent coupling channels. This dimension does not depend on the number of distinct eigenvalues of the coupling operators. The system size only enters in the final boundary layer of the network, making large system sizes accessible to TEMPO methods. 
This structure was discovered by Claude Opus 5.5 during guided algorithm development.
\end{abstract}\maketitle

\paragraph{Introduction}
The time-evolving matrix product operator (TEMPO) method is an efficient algorithm for computing numerically exact dynamics of non-Markovian open quantum systems \cite{strathearnEfficientNonMarkovianQuantum2018, jorgensenExploitingCausalTensor2019, linkOpenQuantumSystem2024a, cochinEfficientConstructionTimeinvariant2026}. The approach relies on a representation of the time-discretized Feynman-Vernon influence functional in terms of a two-dimensional tensor network, spanning a real-time and memory-time dimension \cite{keelingProcessTensorApproaches2026,link2026Tensora}. All variants of the TEMPO algorithm use essentially the same network structure, and differ only in the way the network is contracted in order to compute dynamics \cite{richter2026Numerically}. A drawback of this approach is that the dimensions of the internal network bonds scale quadratically with the dimension of the coupled system Hilbert space and exponentially with the number of coupling channels \cite{cochinEfficientConstructionTimeinvariant2026, link2026Tensora}. Here we show that the bulk network dimensions can be made independent of the system Hilbert space size and only quadratic in the number of coupling channels. The reduction arises through a low-rank structure in the network gates that has previously not been exploited analytically. We illustrate the direct numerical advantage through a benchmark simulation of a damped bosonic mode and present simulation results for a challenging multi-channel problem that was previously not tractable with TEMPO.

\begin{figure}[t]
\centering
\includegraphics{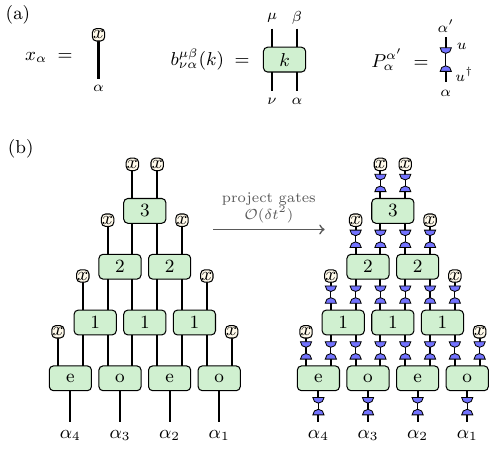} 
\caption{(a) Graphical notation for the boundary vector $x$, the gate $b(k)$ and the low-rank projector $P=uu^T$. Bold legs describe indices with dimension $d^{2m}$ whereas thin legs have dimension $(2m+1)(m+1)$. (b) TEMPO network as in Ref.~\cite{link2026Tensora}, for $N=4$ time steps. Applying the projector to all network bonds induces an error of $\mathcal{O}(\dt^2)$.}
\label{fig:network}
\end{figure}

\paragraph{TEMPO network gates}
We consider the standard open system model $H(t)=H_\mathrm{sys}\otimes \id_\mathrm{env}+\sum_{l=1}^m S_l\otimes B_l(t)$ with Hermitian coupling operators $S_l$ and a Gaussian bosonic bath with correlations $C_{ll'}(t-t')=\langle B_l(t)B_{l'}(t')\rangle$ \cite{link2026Tensora}. Non-Hermitian couplings $L\otimes B^\dagger+L^\dagger\otimes B$ are
written with two Hermitian operators such as $S_1=L+L^\dagger$, $S_2=\ii (L^\dagger-L)$. With a fixed small time step $\dt$ and discretized
correlations $\eta_k\approx (\delta t)^2 C(k\,\delta t)$ ($m\times m$ matrix), the influence functional can be written as a two-dimensional tensor network depicted in Fig.~\ref{fig:network}, see Ref.~\cite{link2026Tensora} for a detailed derivation. This network is composed of elementary gates containing the tensor
\begin{equation}
  I_{\alpha\beta}(k)=\exp\!\Big(-(\bm s^{+}_\beta-\bm s^{-}_\beta)^T\big(\eta_k\,\bm s^{+}_\alpha-\eta_k^{*}\,\bm s^{-}_\alpha\big)\Big),
  \label{eq:gate}
\end{equation}
where $\bm s^{\pm}_\alpha\in\mathbb R^m$ are eigenvalues of the coupling operators on the forward ($+$) and backward ($-$) Keldysh contour, labeled by a multi-index $\alpha$. In general this index has the dimension $d^{2m}$, where $d$ is the system size and $m$ is the number of channels. We continue with the simplest case $m=1$ to demonstrate the procedure for reducing the index dimensions to exactly 6 for this case. We first group all monomials of eigenvalues up to second order in a vector $((f_{\alpha})_n)=(1, s^+_\alpha,s^-_\alpha, (s_\alpha^+)^2,s^+_\alpha s^-_\alpha ,(s_\alpha^-)^2 )$, with the vector index $n=1\ldots 6$. Then we expand the exponential function in $I(k)$ to second order, which yields a quadratic form in $f$
\begin{equation}
\begin{aligned}
I_{\alpha\beta}(k)=\sum_{n,n'=1}^{6}(f_\alpha)_n\,D_{nn'}(k)\,(f_\beta)_{n'}+\mathcal O(\dt^6),
\end{aligned}
  \label{eq:gate2}
\end{equation}
with the $6\times 6$ coefficient matrix ($\eta=\eta_k$)
\begin{equation}
  D(k)=\begin{pmatrix}
  1
  \end{pmatrix}\oplus
    \begin{pmatrix}
  -\eta&\eta\\
  \eta^*&-\eta^*
  \end{pmatrix}\oplus
  \begin{pmatrix}
  \tfrac12\eta^2&-\eta^2&\tfrac12\eta^2\\
  -|\eta|^2&2|\eta|^2&-|\eta|^2\\
  \tfrac12\eta^{*2}&-\eta^{*2}&\tfrac12\eta^{*2}
  \end{pmatrix}.
  \label{eq:C}
\end{equation}
Neglecting higher orders of the exponent introduces only a small error of $\mathcal{O}(\dt^6)$. We can simplify the tensor by performing a singular value decomposition (SVD) of the matrix $F_{\alpha n}=(f_\alpha)_n$ which yields the factorization $F=u\sigma v^T$ with isometry $u\in \mathbb{R}^{d^2\times 6}$. In other words, $u$ is an orthonormal basis of the column span of $F$. This reveals the low-rank structure of $I(k)$ explicitly
\begin{equation}\label{eq:I_invar}
I(k)=u\tilde I(k)u^T + \mathcal{O}(\delta t^6)
\end{equation}
with the reduced tensor ($6\times 6$)
\begin{equation}
\tilde I_{mm'}(k)= \sum_{n,n'=1}^6\sigma_{m}v_{nm} D_{nn'}(k) v_{n'm'}\sigma_{m'}.
\end{equation}
The orthogonal projector onto this quadratic polynomial subspace is $P=uu^T$, and Eq.~\eqref{eq:I_invar} implies $I(k)\approx P I(k)P$. Note that strong compressibility of $I(k)$ has been identified numerically in Ref.~\cite{cochinEfficientConstructionTimeinvariant2026}, without invoking an explicit analytical construction. Polynomial representations of the influence functional have also been employed in Ref.~\cite{grimm2026Integral}, using a construction different from the TEMPO bond projection considered here.
\paragraph{Network reduction}
In the TEMPO network $I(k)$ enters through the four-leg tensor $b^{\mu\beta}_{\nu\alpha}(k)=\delta_{\alpha\mu}\delta_{\beta\nu}I_{\alpha\beta}(k)$. Due to the additional $\delta$-functions in this tensor, the approximate invariance under projection present in $I(k)$ is lost. However, the projection $b(k)\rightarrow (P\otimes P) b(k) (P\otimes P)$ still works out in the full TEMPO network up to quadratic errors in the time-step. The TEMPO network for $N=4$ time steps is displayed in Fig.~\ref{fig:network}(b), left side. The boundaries of the network are capped with trivial identity vectors $x_\alpha\equiv 1$. These constant boundaries are order-zero monomials in $\bm s^\pm$, which are included in the projector's subspace $x = Px$. Any higher-order monomial incoming to a network gate $b(k)$ must have encountered previous interactions that each carry a coefficient of order $\dt^2$. 
For projecting the bonds of a single gate at layer $k$, the leading error comes from incoming quadratic monomials with two earlier interactions. There are $N-k$ preceding interaction layers, so that these incoming quadratic monomials sum up to a total weight of order $\mathcal O((N-k)^2\delta t^4)$.
Projection of a single gate therefore comes with a leading accuracy cost of $\mathcal{O}((N-k)^2\delta t^6)$ from neglecting outgoing third-order monomials. At fixed physical time $T=N\delta t$ this error is bounded by $\mathcal{O}(N^2\delta t^6)=\mathcal{O}(\delta t^4)$. Projecting all $\mathcal{O}(N^2)$ network gates onto the second-order monomials, as in Fig.~\ref{fig:network}(b), thus produces an overall error of $\mathcal{O}(N^2\dt^4)=\mathcal{O}(\dt^2)$ for the influence functional. This is the same error order in $\delta t$ as induced by the symmetric Trotter splitting that TEMPO methods carry anyway. 
The projected gates can be written as
\begin{equation}
(P\otimes P)b(k)(P\otimes P)
=
(u\otimes u)\tilde b(k)(u^T\!\otimes u^T)
+\mathcal O(\delta t^6),
\end{equation}
with
\begin{equation}
\tilde b^{ab}_{cd}(k)
=
\sum_{p,q} T_{adp}\,\tilde I_{pq}(k)\,T_{bcq},
\end{equation}
and $T_{abc}=\sum_\alpha u_{\alpha a}u_{\alpha b}u_{\alpha c}$.
The projected network then takes the same form as the original network but with the reduced gates $\tilde b(k)$ and $\tilde x=u^Tx$. The dimension of all network bonds then becomes $6$ rather than $d^2$, i.e.~independent of the system dimension (more precisely, $d$ is the number of distinct eigenvalues of the coupling operator $S$). This procedure generalizes directly to the case of multiple channels $m>1$. Then the network dimension is reduced to the number of monomials built from all eigenvalues $\bm s^{\pm}_\alpha$
up to degree 2, in total $(2m+1)(m+1)$ monomials. This represents the upper limit of the reduction, linear dependencies among the evaluated monomials or additional structure of the bath correlations can reduce it further.

\begin{figure}[t]
\centering
\includegraphics[width=0.8\columnwidth]{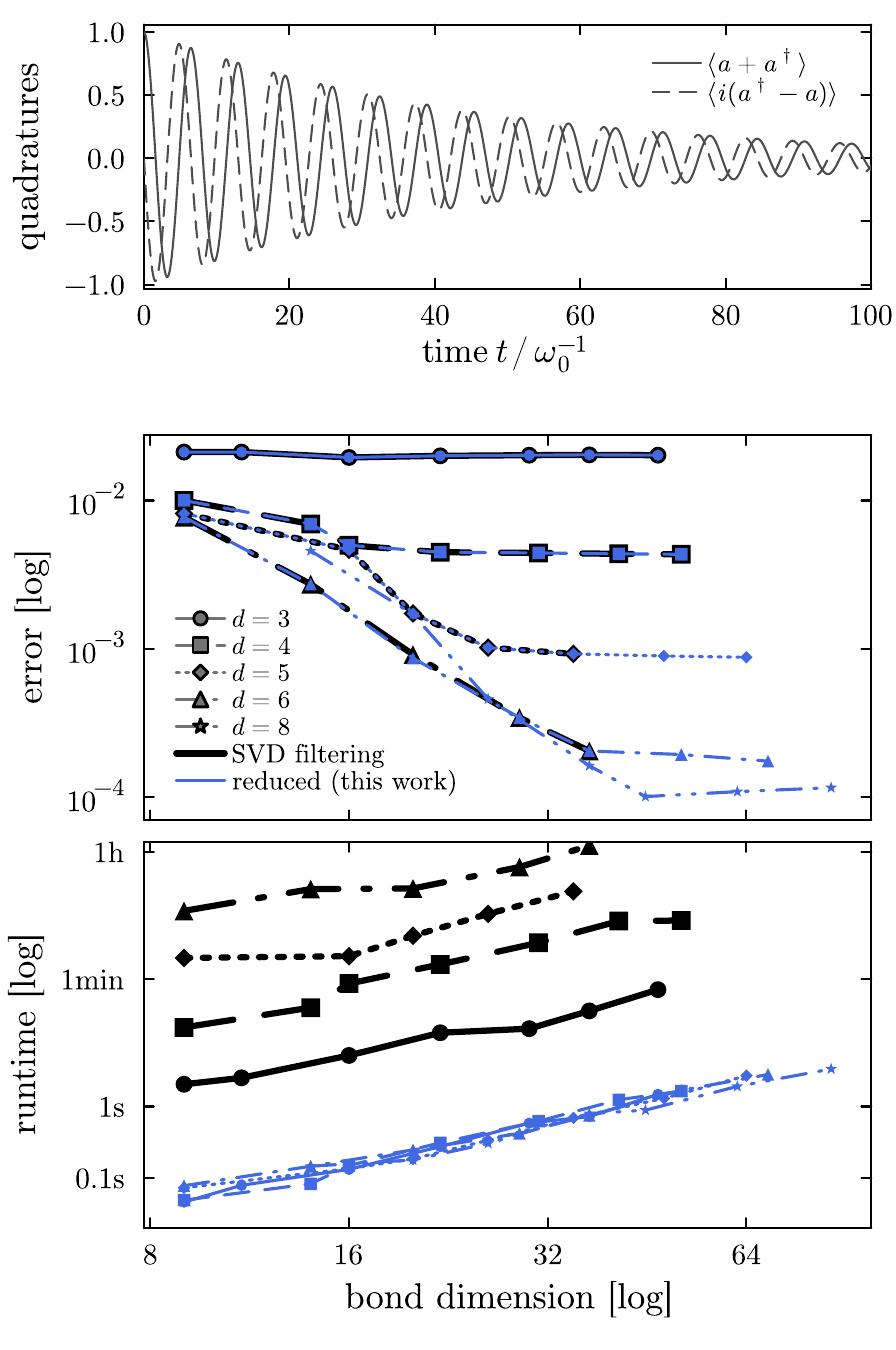}
\caption{Benchmark of uniTEMPO for the model \eqref{eq:DHO} ($\eta=0.01$, $\omega_c=3\omega_0$) using the reduced network compared to using the full network and SVD filtering (Ref.~\cite{cochinEfficientConstructionTimeinvariant2026}). Top: Example dynamics used to quantify the errors. Center: Error of the dynamics measured by the maximum trace norm of the difference between the time-evolved system state and the exact reference. At low $d$ errors are dominated by the Fock-space cutoff. Bottom: Runtimes reported for contracting the influence functional to MPOs with different bond dimensions for $\dt = 0.025/\omega_0$ on an AMD EPYC 7313 CPU (one Julia thread, two OpenBLAS threads, Julia 1.11). }
\label{fig:benchmark}
\end{figure}

\paragraph{Benchmark} As a benchmark example we consider the damped harmonic oscillator
\begin{equation}\label{eq:DHO}
H(t)=\omega_0 a^\dagger a + a^\dagger B(t) + a B^\dagger(t)
\end{equation}
with a zero-temperature Gaussian Ohmic bath 
\begin{equation}
C(t-t')=\braket{B(t)B^\dagger(t')}=\int_0^\infty J(\omega)\eul^{-\ii\omega (t-t')}\diff \omega,
\end{equation}
and $J(\omega)=\eta \omega \exp(-\omega/\omega_c)$. In the original TEMPO network for this model all bonds scale as $d^4$, where $d$ is the maximum number of Fock levels taken into account for the system. In the reduced network the bonds only depend on the number of independent coupling channels. For the zero-temperature bath there is only one independent channel, $m = 1$, because $\braket{B^\dagger(t)B(t')}=0$ makes the correlation matrix rank one. For contracting the network we use uniTEMPO \cite{linkOpenQuantumSystem2024a}, which corresponds to a contraction in reverse memory time, i.e.~from top to bottom in Fig.~\ref{fig:network}(b). This yields a matrix product operator (MPO) form of the influence functional with a homogeneous bond dimension representing the compressed environment. As can be seen in the lower panel of Fig.~\ref{fig:benchmark}, the runtime for the contraction of the new network is independent of $d$, whereas the filtering method from Ref.~\cite{cochinEfficientConstructionTimeinvariant2026} based on the non-reduced network becomes expensive at large $d$.
In order to quantify the error of both methods we consider quench dynamics starting with an initial coherent state, displayed in the top panel of Fig.~\ref{fig:benchmark}. We measure the maximum trace norm deviation with respect to an exact Gaussian reference that was truncated to the corresponding $d$-dimensional Fock space. Both methods give the same accuracy at fixed bond dimension. At low $d$ the accuracy is limited by the Fock-space truncation.

\begin{figure}[t]
\centering
\includegraphics[width=0.8\columnwidth]{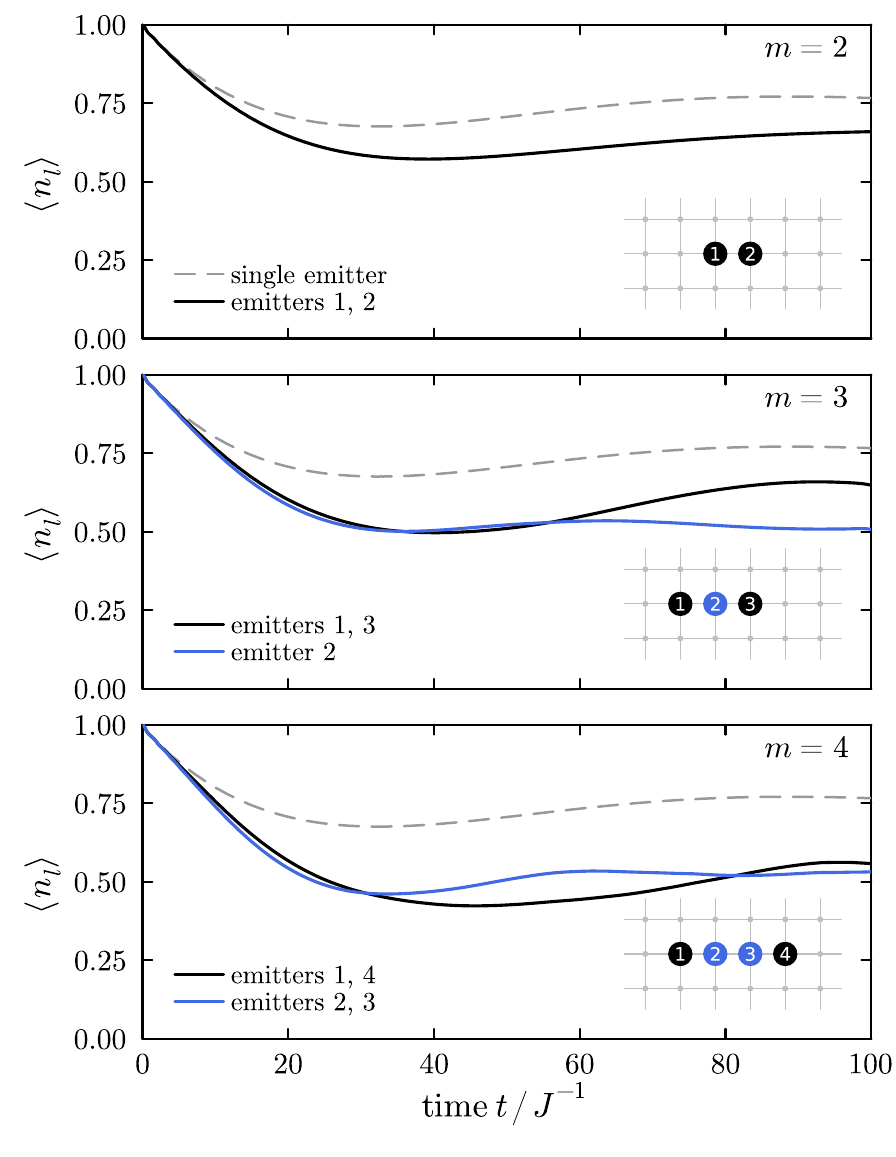}
\caption{Relaxation dynamics of the emitter occupation $\braket{n_l}=\braket{\sigma^+_l\sigma_l^-}$ for the model \eqref{eq:emitter_model1} with $\omega_\mathrm{e}=-4J$ and $g^2=0.1J^2$. The coupling geometry is illustrated in the insets, dashed lines show the relaxation of a single emitter. Simulations were performed using the causal TEMPO algorithm \cite{richter2026Numerically} at $N_c=N$, $\delta t=0.1/J$ with the reduced network and SVD truncation tolerances $4\times 10^{-8}$ ($m=2$), $1.2\times 10^{-6}$ ($m=3$) and $4\times 10^{-5}$ ($m=4$). }
\label{fig:emitter}
\end{figure}

\paragraph{Multi-channel example} To demonstrate the capabilities of the approach for simulating multi-channel systems, we consider a model of multiple emitters coupled to a shared structured reservoir~\cite{Gonzales2017,papaefstathiouEfficientTensornetworkSimulation2025,link2026Tensora}. In particular, $m$ two-level emitters are coupled locally to a two-dimensional bosonic lattice at different lattice sites $\vec{x}_{l}$. Introducing local bosonic bath modes $[b_{\vec{x}},b_{\vec{y}}^\dagger]=\delta_{\vec{x},\vec{y}}$ the full Hamiltonian reads
\begin{equation}\label{eq:emitter_model1}
\begin{split}
    H=&\sum_l \frac{\omega_\mathrm{e}}{2} \sigma_l^z+ g\sum_{l=1}^m (\sigma^-_l b_{\vec{x}_l}^\dagger + \sigma^+_lb_{\vec{x}_l})\\&-J\sum_{\langle \vec{x},\vec{y}\rangle} (b_{\vec{x}}^\dagger b_{\vec{y}} + b_{\vec{y}}^\dagger b_{\vec{x}}),
\end{split}
\end{equation}
where $\langle \vec{x},\vec{y}\rangle$ denotes a sum over nearest neighbors in a square lattice and $\sigma^{z,\pm}_l$ act on emitter $l$. We assume that the bath is at zero temperature. Previous TEMPO methods were restricted to $m=2$ for this system due to the large network bonds of dimension $2^{4m}$  \cite{link2026Tensora}, versus $(2m+1)(m+1)$ for the reduced network. In the example simulation presented in Fig.~\ref{fig:emitter}, emitters are placed on nearest-neighbor lattice sites in a row ($\vec{x}_l=la\vec{e}_x$ with lattice spacing $a$). All emitters are initially in their excited state, so that the system carries $m$ excitations in total. We set the emitters resonant with the band edge of the bosonic lattice $\omega_\mathrm{e}=-4J$. Then the interaction is dominated by low-momentum bath modes, leading to strongly non-Markovian decay with partial trapping of excitations in a localized bound state. When the number of emitters is increased, the decay is accelerated by collective emitter relaxation \cite{Gonzales2017,papaefstathiouEfficientTensornetworkSimulation2025}. For simulating this problem we used the reduced network in combination with the causal TEMPO contraction scheme suggested in Ref.~\cite{richter2026Numerically}. The simulations for $m=3$ and $m=4$ with bond dimension around $200$ took about two days on a single CPU.
\paragraph{Conclusion} The dimensional reduction presented in this work makes it possible to directly apply TEMPO methods to larger systems that were previously not accessible with this approach. This new structure can serve as a conceptual starting point for further generalizing the method, for instance to support time-dependent coupling operators or coupling operators with a continuous spectrum.

\paragraph{AI statement}
This result was discovered by Claude Opus 5.5 during guided development of TEMPO methods. AI-generated code, cross-checked against our established code base, is available under Ref.~\cite{github_repo}.

\bibliography{bib}

\end{document}